\documentclass[submission,copyright,creativecommons]{eptcs}
\usepackage{amssymb}
\usepackage{adjustbox}
 \newcommand{\rimp}{\Rightarrow}
\newcommand{\dimp}{\Leftrightarrow}
\usepackage{float}
\floatstyle{boxed} 
\restylefloat{figure}
\usepackage{wrapfig}
\newfloat{figure}{h}{ext}
\floatname{figure}{Figure}
\usepackage{graphicx}
\usepackage{epstopdf}

\usepackage{algorithmic}
\floatstyle{boxed}
\newfloat{algorithm}{htbp}{loa}
\floatname{algorithm}{Algorithm}

\newcommand{\be}{\begin{enumerate}}
\newcommand{\ee}{\end{enumerate}}

\newcommand{ \outcome}{{\tt outcome}}
\newcommand{ \decision}{{\tt decision}}
\newcommand{\abort}{{\it abort}}

\newcommand{\false}{{\it false}}

\newcommand{\no}{{\it no}}
\newcommand{\yes}{{\it yes}}
\newcommand{\commit}{{\it commit}}

\newcommand{\undecided}{{\it undecided}}
\newcommand{\undef}{{\it undef}}

\newcommand{\Prop}{{\it Prop}}

\newcommand{\vote}{{\tt vote}}

\newcommand{\ack}{{\tt ack}}

\newcommand{\I}{{\cal I}}
\newcommand{\R}{{\cal R}}

\newcommand{\Next}{X}

\newcommand{\powerset}[1]{{\cal P}(#1)}

\newcommand{\nat}{{\bf N}}

\title  {Epistemic Model Checking of Atomic Commitment Protocols with Byzantine faults\thanks{Thanks to Ron van der Meyden for some preliminary discussions on the topic of this paper.}}

\author   {Omar Ibrahim Al-Bataineh \\ School of Computer Science  \\ University of Liverpool, UK}

\date{}

\begin{document}

\maketitle

\begin{abstract}

The notion of knowledge-based program introduced by Halpern and Fagin provides a useful formalism for designing, analysing, and optimising distributed systems. This paper formulates the two phase
commit protocol as a knowledge-based program and then
an iterative process of model checking and counter-example guided refinement is followed
to find concrete implementations of the program for the case of perfect recall semantics in the Byzantine faults context with synchronous reliable communication. 
We model several different kinds of Byzantine faults and verify different strategies to 
detect them. We address a number of questions that have not been considered in the prior literature, viz.,  under what circumstances a sender can know that its transmission has been successful,
and under what circumstances an agent can know that the coordinator is cheating, and find
concrete answers to these questions.
The paper also describes a methodology based on temporal-epistemic model checking technology 
that can be followed to verify the shortest and longest execution time of a distributed protocol 
and the scenarios that lead to them. 
\end{abstract}

\section{Introduction}

Distributed systems have proven to be hard to understand, design, and reason
about due to their complexity and non-deterministic nature.
They usually involve subtle interactions of a number of components in contexts of unpredictable
environment. Reasoning about actions in distributed systems at the level of knowledge allows us to abstract away
many concrete details about the systems which helps to 
simplify analysis of distributed systems.
In particular, the notion of knowledge-based program introduced by Halpern and Fagin \cite{FHMVbook}
provides an abstract level of description of systems in which an agent's actions depend explicitly on the agent's knowledge.

Knowledge-based programming has several advantages over classical programming 
(i.e. a style of programming that does not  explicitly use knowledge-based tests).
First, knowledge-based programming can be used to build an optimal solution
for the given problem in the sense that agents do not overlook opportunities 
to use relevant information that is available in their local states.
Second, this style of programming allows us to modify the
developed program more easily when considering a context with different properties
such as different failure assumptions or different communication topology.

Unlike standard programs, knowledge-based programs cannot be
directly executed, since the satisfaction of the knowledge subformulas in the knowledge-based program
depends on the set of all runs of the program, which depends on the
actions taken, which in turn depends on the satisfaction of these knowledge
subformulas. It has been proposed to address this apparent circularity by treating knowledge-based programs
as specifications, and defining when a concrete standard program satisfies this
specification \cite{FHMVbook}. However, due to the high complexity of information flows in distributed systems, 
finding a concrete implementation for a given knowledge-based program can be a nontrivial task.


In this work we use the partially automated methodology proposed in \cite{AlBatainehMeyden10,OmarME2011} to find a concrete implementation for knowledge-based programs (KBPs) for the case of \textit{perfect recall semantics} in which agents make optimal use of their observations. The methodology is based on an iterative process of model checking and counterexample guided refinement that  helps the user to find predicates on the agent's observable variables that correspond \textit{precisely} to the knowledge conditions given in the KBP.  We use the epistemic model checker MCK \cite{mck}, to develop concrete implementations of knowledge-based programs for extensions of the basic two phase commit protocol (2PC) \cite{Ber87}.

Atomic commitment, of which the 2PC protocol is an example, is a problem of central importance to distributed  databases \cite{Hadzilacos87,Mazer90} and the currently more popular atomic transactions tasks. 
The protocol can be used in many interesting contexts and it has several applications in databases and computer networking. As is often the case, in concurrency and distributed computing,
even minor failures make a thorough understanding and the development of correct solutions a very subtle issue. In this work, we extend the known solutions by considering additional forms of faults, as well as by analysing questions that become available within the epistemic framework, such as ``when does a process know that the controller is cheating?", and ``can all cases of cheating be detected?"

The objective of the new extension that we consider is to integrate a ``cheating detection'' mechanism with the 2PC protocol
while allowing the controller to behave maliciously.  
The goal is then to derive an implementation of the protocol that is able to understand when the controller is cheating
and to enable the agents to respond appropriately when a cheating is detected. The perfect recall semantics of knowledge is particularly relevant for the analysis of atomic commitment with Byzantine faults,
where optimal use of information is of concern. 
This issue is significant in the extension that we study in particular 
when we allow the controller to cheat from one round to another.
We also describe  a methodology based on temporal model checking that can be followed
to find the lower and upper bound for termination in each of the contexts that we study.

\paragraph{Related Work.}

The various failure modes in distributed systems have received great attention from the researchers in the last two decades. 
The literature has concentrated mainly on three basic failure modes: crash failures, omission failures, and Byzantine failures. However, robustness of distributed systems against Byzantine
attacks have been of prime importance and an active field of research for many years. Byzantine behaviour also is subject to intensive research in computer security and cryptography. 
For instance, there is a large body of work in the area of secure multi-party computation \cite{Yao1982}.
Byzantine systems will be increasingly important in the future since malicious attacks are increasingly common and can cause faulty nodes to act in any arbitrary manner.

Several researchers have used the theory of process knowledge 
in design and analysis of distributed agreement protocols, of which atomic commit protocols are examples of these protocols. The works in \cite{Hadzilacos87,Mazer90} give a knowledge-theoretic treatment of atomic commitment. 
The authors show the minimum knowledge levels that hold in two phase and three phase atomic commitment protocols.
They determine what knowledge processes need to commit or abort a transaction
under the assumption that processes can crash and recover. However, the approach described in their work is a manual approach and the analysis is carried out by pencil and paper reasoning. The authors also have not studied the protocol in a Byzantine faults context.


Knowledge-based programs have been used in papers such as \cite{DM90,HZ92} in
order to help in the design of new protocols or to clarify the understanding of existing
protocols. Examples of the development of standard programs from knowledge-based
programs can be found in \cite{DM90,SR86}. The approach described in these papers
is different from the one we discussed here in that it is done by pencil and paper
analysis using the theory of knowledge.
Examples of the use of epistemic model checkers to identify
implementations of knowledge-based programs remain limited.
 One is the work of Baukus and van der Meyden \cite{BaukusM04} who use MCK to analyse several protocols for the cache coherence problem using knowledge-based framework.
Another work is the work of Al-Bataineh and van der Meyden \cite{AlBatainehMeyden10} 
who use MCK to analyse the Herbivore protocol, an anonymous broadcast protocol, using knowledge-based programs. 

Recently,  Huang and van der Meyden \cite{HuangM14}  develop an approach that automates the construction of implementations for knowledge-based programs for the case of the observational semantics for knowledge-based programs.  However, such semantics are not appropriate for the example we study in this paper where we would like to derive implementations of knowledge-based programs of the 2PC protocol while allowing agent to recall their past observations.

\section{Epistemic Logic and MCK}

Epistemic logics are a class of modal logics that
include operators whose meaning concerns
the information available to agents in a 
distributed or multi-agent system. 
We describe here briefly a
version of such a logic combining operators for 
knowledge and linear time, and its semantics in 
a class of structures  known in the literature as  {\em interpreted 
systems} \cite{FHMVbook}. 
We then discuss the model checker MCK \cite{mck}, which is based on this semantics. 

Suppose that we are interested in systems comprised of $n$ agents
and a set $\Prop$ of atomic propositions. 
The syntax of the fragment of the logic of knowledge and time relevant for this paper 
is given by the following grammar: 
$$
\phi ::= \top \mid p \mid \neg \phi \mid \phi \wedge \phi \mid K_i\phi \mid \Next\phi \mid G \phi \mid F \phi
$$
where $p\in \Prop$ is an atomic proposition and $i\in \{1\ldots n\}$ is an agent. 
(We freely use standard boolean operators that can be defined 
using the two given.) Intuitively, the meaning of $K_i\phi$ 
is that agent $i$ knows that $\phi$ is true, $\Next\phi$ means that 
$\phi$ will be true at the next moment of time, $G \phi$ means
that $\phi$ holds now and all future time, and $F \phi$ means in the future (or now) $\phi$ holds.

The semantics we use is the  {\em interpreted systems} model 
for the logic of knowledge \cite{FHMVbook}. For each $i=0\ldots n$, 
let $S_i$ be a set of states. For $i=0$, we interpret $S_i$ 
as the set of possible states of the environment within which the agents operate; 
for $i=1\ldots n$ we interpret $S_i$ as the set of {\em local states} of agent $i$. 
Intuitively, a local state captures all the concrete pieces of information on the 
basis of which an agent determines what it knows. 
We define the set of {\em global states} based on such a collection of environment and 
local states, to be the set $S=S_0\times S_1\times \ldots \times S_n$. 
We write $s_i$ for the $i$-th component (counting from 0) of a global state $s$. 
A {\em run} over $S$ is a function $r:\nat \rightarrow S$. 
An {\em interpreted system} for $n$ agents is a tuple $\I = (\R,  \pi)$, 
where $\R$ is a set of runs over $S$, and $\pi: S\rightarrow \powerset{\Prop}$ is an interpretation function. 

A {\em point} of $\I$ is a pair $(r,m)$ where $r\in \R$ and $m\in \nat$. 
We say that two points  $(r,m),(r',m')$ are {\em indistinguishable} to agent $i$, and write $(r,m)\sim_i(r',m')$, 
if $r(m)_i = r'(m')_i$, 
i.e., if agent $i$ has the same local state at these two points. 
We define the semantics of the logic by means of a relation $\I,(r,m)\models \phi$, 
where $\I$ is an intepreted system, $(r,m)$ is a point of $\I$ and $\phi$ is a formula. 
This relation is defined inductively as
follows: 
\begin{itemize} 
\item $\I,(r,m) \models p$ if   $p\in \pi(r(m))$,  
\item 
$\I,(r,m)\models \neg \phi$ if not $\I,(r,m)\models \phi$,
\item 
$\I,(r,m)\models \phi_1\lor \phi_2$ if $\I,(r,m)\models \phi_1$ or $\I,(r,m)\models \phi_2$ ,
\item 
$\I,(r,m)\models \Next \phi$ if $\I,(r,m+1)\models \phi$,
\item
$\I,(r,m)\models  G \phi$ if for all $m' \geq m$ we have $ \I,(r,m') \models \phi$,
\item
$\I,(r,m)\models  F \phi$ if there is $m' \geq m$ such that $ \I,(r,m') \models \phi$,
\item 
$\I,(r,m)\models K_{i} \phi$ if  for all  points $(r',m')$ of $\I$  such that $r_{i}(m)= r'_{i}(m')$  we have $\I,(r',m') \models \phi$.
\end{itemize} 

MCK \cite{mck} is a model checker based on this semantics for the logic of knowledge. 
For a given interpreted system $\I$, and a specification $\phi$ in the  logic of knowledge and time, MCK computes whether 
$\I,(r,0) \models \phi$ holds for all runs $r$ of $\I$.  
Since interpreted systems are infinite structures,
MCK  allows an interpreted system to be given a finite description in the 
form of a program from which the interpreted system can be generated. 
It supports different semantics of knowledge, however, in this paper we will work with respect to perfect recall semantics of knowledge in which agents make optimal use of their observations, 
by taking all observations made in the past into account when computing what they know.
This issue is significant in the example that we study in this paper.

\section{Knowlege-based Programs} \label{sec:methodology}

Knowledge-based programs \cite{FHMVbook} are like standard programs, except that 
expressions may refer to an agent's knowledge.
That is, in a knowledge-based  program for agent $i$, we may find statements  of the form  ``$v:= \phi$", 
where $\phi$ is a formula of the logic of knowledge.
In general, knowledge-based programs may have no implementations, a behaviourally unique implementation, or
many implementations. Some conditions are known under which a behaviourally unique implementation is 
guaranteed to exist. One of these conditions is that agents have perfect recall and 
all knowledge formulas in the program refer to the present time  \cite{Meyden1996}. 
This case will apply to the knowledge-based programs we consider in this paper.
Suppose that we have a standard program $P$ of the same syntactic structure as the knowledge-based program ${\bf P}$, 
in which each knowledge-based expression 
$\phi$ is replaced by a concrete predicate $p_\phi$ of the observable variables of the agent. 
In order to handle the perfect recall semantics, we  
also allow $P$ to add local {\em  history variables} and code fragments of the form $v:=e$, where $e$ is an expression, 
that update these history variables, so as to make information about past states available at the
current time.  

The concrete program $P$ generates a set of runs that we can take to be the basis of an interpreted system 
$\I(P)$. We now say that $P$ is an {\em implementation} of the knowledge-based program ${\bf P}$ if
for each formula $\phi$ in a conditional, we have that in the interpreted system 
$\I(P)$.  The formula $p_\phi \dimp \phi$ is valid (at times when the condition is used).
That is, the concrete condition is equivalent to the knowledge condition in the implementation. 
 so we are guaranteed behaviourally unique implementations. 

We now describe a partially automated process, using epistemic model checking, 
that can be followed to find implementations of knowledge-based programs \cite{AlBatainehMeyden10,OmarME2011}.
The user begins by introducing a local boolean variable $v_\phi$ for each knowledge formula
$\phi= K_i\psi$ in the knowledge-based program, and replacing $\phi$ by $v_\phi$. 
Treating $v_\phi$  as a history variable, the user may also add to the program statements of the form $v_\phi:=e$, 
relying on their intuitions concerning situations under which the epistemic formula $\phi$ will be true. 
This produces a standard program $P$ that is a candidate to be an implementation of the knowledge-based program
${\bf P}$. 

To verify  the correctness of $P$ as an implementation of ${\bf P}$, the user must now 
check that the variables $v_\phi$ are being maintained so as to be equivalent to the 
knowledge formulas that they are intended to express. 
This can be done using epistemic model checking, where we verify formulas of the form 
$$X^n (pc_i=l \rimp (v_\phi \dimp K_i\psi) )$$
where $n$ is a time at which the test containing $\phi$ may be executed, 
$pc_i$ is the program counter of agent $i$ and $l$ is a label for the 
location of the 
expression
containing $\phi$. 
In general, the user's guess concerning the concrete condition that is equivalent to the  knowledge formula 
may be incorrect, and the model checker will report the error.
The next step of our process requires the user to analyse this error trace
(by inspection and human reasoning) in order to understand the source of the error in their guess for the concrete condition representing the knowledge formula.
As a result of this analysis, a correction of the assignment(s) to the variable $v_{\phi}$ is
made by the user (this step may require some ingenuity on the part of the user.)
The model checker is then invoked again to check the new guess. This process is
iterated until a guess is produced for which all the formulas of interest are found
to be true, at which point an implementation of the knowledge-based program
has been found.

\section{Atomic Commitment Protocols}

The atomic commitment protocols (ACPs) are a class of distributed protocols that aim to coordinate a set of
processes that participate in a distributed transaction on
whether to commit or abort (roll back) the transaction,
while preserving the consistency of the distributed databases.
An atomic commitment protocol for distributed transactions is 
the one that satisfies the following requirements \cite{Ber87}:
\begin{itemize}

\item \textbf{R1}. (Atomicity): All-or-nothing, a transaction is committed if and only if all sub-transactions it depends on also commit.

\item \textbf{R2}. (Consistency): All non-faulty processes decide on the same value.

\item \textbf{R3}. (Validity): The validity requirement can be classified into two requirements: the \textit{commit-validity} requirement which states that if all processes vote `yes', then the decision will be commit, and the \textit{abort-validity} requirement which states if there is a vote saying `no',  then the decision will be to abort.
 
\end{itemize}

We now give an informal description of the 2PC protocol that solves
the problem of processing a distributed transaction in a synchronous setting \cite{Ber87}.

A set of processes $\{p_{1},.., p_{n}\}$ participate in a distributed transaction. 
Each process has been given its own subtransaction.
One of the processes will act as a coordinator and all other
processes are participants.
The protocol proceeds into two phases.
In the first phase (voting phase),
the coordinator broadcasts a start message
to all the participants, and
then waits to receive vote messages from the participants.
The participant will vote to commit the transaction if all its local computations
regarding the transaction have been completed successfully;
otherwise, it will vote to abort.
In the second phase (commit phase), if the coordinator
received the votes of all the participants, 
it decides and broadcasts the decision.
If all the votes are `yes' then the coordinator will
decide to commit the result of the transaction. However, if one
vote said `no', then the coordinator will decide to abort the transaction.
After sending the decision, the coordinator waits to receive a COMPLETION message from all the participants.
The protocol terminates whenever the coordinator receives all COMPLETION messages successfully.

To our knowledge, the protocol has not been verified using model checking in contexts where Byzantine faults are possible. 
In the context of commit protocols, a Byzantine coordinator may process
requests incorrectly 
and/or produce incorrect or inconsistent outputs. 
A crash failure is normally detected by the absence of an expected message within a specified time-out.
A Byzantine fault is much harder than the fail-stop model or the fail-recovery model.
A number of interesting questions  arise when considering a distributed commit protocol in a Byzantine faults context. 
For example, what exact test needs to be applied to determine whether the coordinator is cheating? 
Which participants are able to detect if the coordinator is cheating? 
Are there situations where some participant
is able to detect that the coordinator is cheating and decides to abort the transaction, while some other participant commits? Can the participants figure out whether the coordinator is cheating when increasing 
their observation capabilities using the following means:
(1)  by allowing them to use traps (i.e. to open randomly some runs of the protocol, and allowing the participants to observe each other's vote and decision),  
or (2)  by allowing the agent that knows that the coordinator is cheating to publish its knowledge.

\section{The Two Phase Commit Protocol As Knowledge-based Programs} \label{description}

In this section we give a formulation of the 2PC protocol (see Section 4) as a knowledge-based program.

In Figure \ref{fig:kbp1} we give the knowledge-based program of the coordinator 
and in Figure \ref{fig:kbp2} we give the knowledge-based program of the participant.
For the coordinator's program, the variable \verb+vote[i]+ is used to represent the vote that participant $i$
sends to the coordinator which has three possible values  $\{ \undef, \no, \yes \}$.
Note that if by the end of the first phase $\verb+vote[i]+=\undef$ it 
means that the coordinator has not received the vote of participant $i$ due to a communication failure.
The boolean variable \verb+ack[i]+ is used to record whether $i$ has sent an acknowledgement message to $c$ upon receiving an abort/commit message from it.
There are five environment variables \verb+byzantine+, \decision, \verb+RcvdStartMsg[i]+, \verb+trap+, and \verb+cheatingDetected+.
The boolean variable \verb+RcvdStartMsg[i]+ holds when the environment delivers successfully the start message
from the coordinator to the participant $i$.
The variable \verb+decision+ represents the decision that the coordinator makes during the execution of the protocol
which has three possible values $\{ \undecided, \abort, \commit \}$.
Initially it has the value \textit{undecided}.
The boolean variable \verb+trap+ is observable only by the participants and can be used
by the environment to instruct the participants to open the current run of the protocol.
Similarly, the boolean variable \verb+cheatingDetected+ is observable only by the participants and can be used
by the participant that knows that the coordinator is cheating to publish its knowledge. 
As in message passing systems, neither $c$ nor $i$ takes a \verb+receive+ action; we leave message delivery to the environment's protocol.
The boolean variable \verb+byzantine+ is used to represent whether the agent has a Byzantine behaviour.
The double colon notation `\verb+::+' is used to model non-determinism in the KB programs.

Agents have available the following actions. 
Each agent $i$ can perform the action \verb+send+($x, j, i$) which allows agent $i$ to send
some other agent $j$ the message $x$.
In addition to these actions, the coordinator can perform the action \verb+sendall+$(x)$ 
which allows the coordinator to send the message $x$ to all processes.
We chose message delivery to be completely under the control of the environment. 
This explains why there are no receive actions in the programs. 
We assume also that communication is synchronous
so that there is an upper bound imposed on message delivery delay.
The time delay parameters $\tau_{s}, \tau_{v}, \tau_{k}, \tau_{d}$, and $\tau_{f}$ 
used in the programs have been chosen in a way to guarantee
proper synchronisation between processes. 
Note that all synchronisation issues between the processes are handled in the environment's protocol. 
The boolean function \verb+AllVotesReceived()+ returns true if the coordinator received all participants
votes successfully.

\begin{figure} [hp]
  \setlength{\belowcaptionskip}{0pt}
\begin{small}
{\bf protocol} 2pc-coordinator (\verb+byzantine+, \decision)  \\
\{ \\
local variables: \verb+coord_vote+, \verb+vote[2..n]+, \verb+ack[2..n]+; \\
\hspace*{2pt} \verb+sendall(start)+;   \\ 
\hspace*{2pt} \verb+wait+$(\tau_{v})$;	// wait $\tau_{v}$ time units to receive the participants votes. \\
\hspace*{2pt}  $\verb+if+ (\neg \verb+byzantine+)$ \hspace*{5pt}  // if the coordinator is honest \\
\hspace*{5pt} \{ \\ 
\hspace*{10pt} \verb+if+ (\verb+coord_vote+ = \no)  \verb+then+  \decision ~:= \abort; \\
\hspace*{10pt} \verb+if+ (\verb+coord_vote+ = \yes)  \\
\hspace*{12pt} \{ \\
\hspace*{15pt} \verb+for+ (\verb+i+ =2; \verb+i+ $\leq$ \verb+n+; \verb+i+++) \\
\hspace*{18pt} \{ \\
\hspace*{20pt} \verb+if+ ($\verb+vote[i]+ = \no \lor ~ \verb+vote[i]+ = \undef$) \verb+then+  \{ \decision := \abort; ~  \verb+break+; \} \\	
\hspace*{20pt} \verb+if+ ($\verb+i+ = \verb+n+ ~\land$  \verb+vote[i]+ = \yes) \hspace*{2pt} \verb+then+  \decision  := \commit;\\
\hspace*{18pt}  \} \\
\hspace*{12pt}  \} \\
\hspace*{5pt}  \} \\
\hspace*{2pt} \verb+else+  $\verb+if+ ~ (\verb+byzantine+)$  \\
\hspace*{5pt} \{\\
\hspace*{20pt} $\verb+if ::+$ \decision := \commit; \\
\hspace*{30pt} $\verb+ ::+$   \decision~ := \abort;  \\
\hspace*{20pt} \verb+fi+;\\ 
\hspace*{5pt} \} \\
// The announcement phase \\
\hspace*{5pt}  $\verb+if+ (\neg \verb+byzantine+) \hspace*{5pt}  \verb+then+$ \verb+sendall+ (\decision);  \\
// A faulty coordinator may behave maliciously in any arbitrary way. \\
\hspace*{5 pt}  \verb+else+  $\verb+if+ (\verb+byzantine+)$ \\
\hspace*{10 pt}  \{ \\
\hspace*{15 pt} $\verb+if ::+$ \hspace*{5pt} \verb+sendall+ (\decision);  \\
\hspace*{25 pt}  $\verb+ ::+$ \hspace*{5pt} \{\verb+send+ ($\abort$,~2,~$c$) ; ~~~~ .... ~ ; \verb+send+ ($\commit$,~$n$,~$c$);\}  \\
\hspace*{25 pt} $\verb+ ::+$ \hspace*{5pt} \{\verb+send+ ($\commit$,~2,~$c$) ;~~ .... ~ ; \verb+send+ ($\abort$,~$n$,~$c$);  \}  \\
\hspace*{150 pt}   .... \\
\hspace*{15 pt} \verb+fi+; \\
\hspace*{10pt} \} \\
// Retransmission and termination phase \\
\hspace*{5pt} \verb+while+ (true) \verb+do+  \\
\hspace*{8pt} \{ \\
\hspace*{25pt}  \verb+for+ (\verb+i+ =2; \verb+i+ $\leq$ \verb+n+; \verb+i+++)  \\
\hspace*{30pt} \{ \\
\hspace*{35pt} \verb+if+ ($\neg \verb+byzantine+ \land \neg (K_{c} (\hat{K} _{i} (\verb+decision+))))$   \verb+then+  \verb+send+ (\verb+decision+, $i$, $c$); \\
\hspace*{35pt} \verb+else+  $\verb+if+ ~ (\verb+byzantine+ \land \neg (K_{c} (\hat{K} _{i} (\decision))))$ \\
\hspace*{40 pt}  \{ \\
\hspace*{45 pt} $\verb+if ::+$ \hspace*{5pt} \verb+send+ (\decision, $i$, $c$);  \\
\hspace*{55 pt} $\verb+ ::+$ \hspace*{5pt}  \verb+send+ ($\abort$ , $i$, $c$) ; \\
\hspace*{55 pt} $\verb+ ::+$ \hspace*{5pt}   \verb+send+ ($\commit$, $i$,~$c$) ;   \\
\hspace*{45 pt} \verb+fi+; \\
\hspace*{40 pt} \} \\
\hspace*{30pt} \} \\
\hspace*{25pt} \verb+wait+$(\tau_{k})$;	 \\ 
\hspace*{25pt} \verb+if+ ($\bigwedge_{i=2}^{n} ~  (K_{c} ~  (\hat{K} _{i} (\verb+decision+))))$ \verb+then+  \verb+break+; \\
\hspace*{8pt} \} \\
 \}
\textbf{ \caption{ \label{fig:kbp1}The KBP of the coordinator ($2PC_{c}$)}}
\end{small}
\end{figure}

The term $\verb+cheating+_{c}$ in the $2PC_{i}$ program
represents the condition that the coordinator $c$ has a Byzantine behaviour.
We consider here two possible Byzantine behaviours of the coordinator. 
The first one is to send an incorrect decision to the participants after receiving their votes: it may instruct them to commit while the right
decision is abort and vice versa. 
The second is to send contradictory messages to the participants. 
$$    
 \begin{array}[t]{l}
\verb+cheating+_{c} =  ((\bigvee_{i =2..n} (\verb+vote[i]+ = \no) \lor \verb+coord_vote+ = \no) \land  \bigvee_{i=2..n} (\verb+send+ (\commit,~i,~c))) ~ \lor \\ 
~~~~~~~~~~~~~~~~~~~~~~~ ((\bigwedge_{i=2..n} (\verb+vote[i]+ = \yes) \land \verb+coord_vote+ = \yes) \land  \bigvee_{i=2..n} (\verb+send+ (\abort,~i,~c))) ~ \lor \\
~~~~~~~~~~~~~~~~~~~~~~~ (\bigvee_{i, j =2..n, i \neq j}~ (\verb+send+ (\commit,~i,~c)\land \verb+send+ (\abort,~j,~c))))
\end{array}
$$
\begin{figure} [h]
  \setlength{\belowcaptionskip}{0pt}
  \begin{small}
{\bf protocol} 2pc-part (\verb+RcvdStartMsg+, \decision$[i]$,  \verb+trap+, \verb+cheatingDetected+) \\ 
 \{\\
 \hspace*{5pt} local variables: \vote, \ack, \outcome; \\
\hspace*{10pt}  \verb+if+ (\verb+RcvdStartMsg+)   \\
\hspace*{15pt} \{ \\ 
\hspace*{20pt} \verb+send+(\verb+vote+, $c$, $i$); \\
\hspace*{20pt} \verb+wait+$(\tau_{d})$;	// wait $\tau_{d}$ time units to receive the decision of $c$. \\
\hspace*{20pt} // The decision phase augmented with some cheating detection strategies. \\
\hspace*{20pt} \verb+if+ $(K_{i} (\verb+cheating+_{c})$) ~ \verb+then+  \verb+cheatingDetected+ := \textit{true};  \hfill(1) \\
\hspace*{20pt} \verb+if+ (\verb+trap+ $\land$ $\neg$\verb+cheatingDetected+) \verb+then+ ~ Open2PCRun($i$); \\
\hspace*{20pt} \verb+if+ $(K_{i} (\verb+cheating+_{c}) \lor \vote = \no)$ \verb+then+  \outcome := \abort;   \hfill(2) \\
\hspace*{20pt}  \verb+if+ $(\neg K_{i} (\verb+cheating+_{c}$))  \verb+then+  \outcome ~ := $\verb+decision+[i]$;  \hfill(3)\\
\hspace*{20pt} // Transmission and termination phase \\
\hspace*{20pt} \verb+while+ (true) \verb+do+  \\
\hspace*{25pt} \{ \\
\hspace*{30pt} \verb+if+ ($\neg K_{i} (K _{c} (\hat{K}_{i}(\decision[i])))$) \verb+then+ \verb+send+(\verb+ack+, $c$, $i$);  ~  \verb+wait+$(\tau_{f})$; \\ 
\hspace*{30pt} \verb+if+ ($ K_{i} (K _{c} (\hat{K}_{i}(\decision[i]))$))  \verb+then+ \verb+break+; \\
\hspace*{25pt} \} \\
\hspace*{15pt} \} \\
\}
\textbf{ \caption{ \label{fig:kbp2}The KBP of the participants ($2PC_{i}$)}}
\end{small}
\end{figure} 

The coordinator initially chooses its vote (\verb+coord_vote+) non-deterministically and
then it starts the protocol by sending a start message to the participants.
The coordinator starts the protocol by broadcasting a \verb+start+ message to all the participants on the network. 
If by the end of the 1st phase, some participant's vote was `no', or there was a missing vote due to a communication failure 
($\verb+vote[i]+ = \undef$), and the coordinator is honest, then it decides to abort and broadcasts an abort message. 
On the other hand, if all the participants' votes were `yes' and the coordinator is honest, it decides to commit and broadcasts a commit message.
Note also that the coordinator repeatedly sends the commit or abort
command, until it knows that all the participants have received the decision successfully.
We use $\hat{K}_{i} (\verb+decision+)$ in the programs as an abbreviation 
for $K_{i} (\verb+decision+  = \commit ~\lor~ \verb+decision+  = \abort)$.

All the participants run the same knowledge-based protocol $2PC_{i}$, but with different environment variables.
Similar to the model of the coordinator, each participant initially chooses
its vote non-deterministically and then attempts to communicate the vote to the coordinator.
In case the participants detects that the coordinator is cheating then they decide to abort
the transaction and terminate.
Note that it is necessary that the participants do not follow 
the decision of the coordinator immediately upon receiving it since the coordinator might be faulty.
The participants then need to adapt some strategies to verify whether the coordinator is faulty.
One possible strategy is to use \textit{traps}. That is, to allow the participants
to open randomly some runs of the protocol and then to check whether the coordinator is cheating.
However, to achieve this more communication is required before making a decision.
Moreover, some collaboration among the participants is required:
(1) the participant that does not know whether the received decision
is corrupted has then to wait for sometime before it makes a decision,
(2) the participant that knows that the coordinator is cheating has then to publish his knowledge immediately.

\begin{figure} [h]
  \setlength{\belowcaptionskip}{0pt}
{\bf subprotocol} Open2PCRun(\verb+agent+ $i$) \\ 
 \{\\
\hspace*{5pt} \verb+for+ ( $j$ =2;  $j$ $\leq$ \verb+n+ ; $j$++  ) \\
\hspace*{8pt}\{ \\
\hspace*{15pt} \verb+if+ ($j$ $\neq$ $i$)  \{  \verb+send+(\verb+vote+, $j$, $i$); ~~   \verb+send+(\decision, $j$, $i$); \} \\
\hspace*{8pt}\} \\
\}
\textbf{ \caption{ \label{fig:verification} The Open2PCRun sub-protocol}}
\end{figure} 

\subsection{Correctness Conditions} \label{CorrectnessCond}

The first formula of interest
is global consistency  (i.e. all processes must abort or all must commit.) 

\textit{Specification 1a:  The global consistency is always guaranteed.}
$$ 
 \begin{array}[t]{l} 
 G ~ (\bigwedge_{i\neq j}  \neg  (i.\outcome = \abort \land j.\outcome = \commit))
   \end{array}
$$
An important property of the protocol is that a decision made during its execution
must be taken by all processes. 

 \textit{ Specification 1b:  All processes will eventually decide.} 
 $$ 
 \begin{array}[t]{l} 
   F ~ (\bigwedge_{i=2}^{n}~ (i.\outcome = \commit \lor i.\outcome = \abort))  
 \end{array}
$$
The first property states that no two processes
decide on different values. On the other hand,
the second property
states that it is impossible for a process to remain forever undecided.

Next, we want to verify whether the protocol guarantees the validity property in all the contexts that we consider.
We need to verify here two formulas.

\textit{Specification 2a:} \textit{Commit-validity is always guaranteed.} 
$$ 
 \begin{array}[t]{l} 
  G ~ ((\bigwedge_{i=2..n} (\verb+vote[i]+ = \yes) \land  \verb+coord_vote+ = \yes) \rimp   F ~  (\verb+decision+ = \commit))
   \end{array}
$$

\textit{Specification 2b:} \textit{Abort-validity is always guaranteed.} 
$$ 
 \begin{array}[t]{l} 
  G ~ ((\bigvee_{i=2..n} (\verb+vote[i]+ = \no) \lor (\verb+coord_vote+ = \no)) \rimp   F ~ (\verb+decision+ = \abort))
   \end{array}
$$

For the Byzantine-faults context, we want to check whether the participants are able to detect if the coordinator is cheating. 

\textit{Specification 3: If coordinator $c$ is cheating then everyone knows $c$ is cheating.}
$$
X^{f}((\verb+cheating+_{c}) \rimp K_{i} (\verb+cheating+_{c}))
$$
In general, Specification 3 needs to be verified at specific times $f$, straightforwardly determined 
from the structure of the $2PC_{i}$ program. We mark the places at which we need to verify $K_{i} (\verb+cheating+_{c}$) by asterisks 
so that it is easy for us to refer to them  when discussing the implementations of the program.

Finally, we want to ensure that the protocol can always succeed to terminate.
We say that the protocol has successfully terminated if the coordinator and all the participants
have terminated. Note that the coordinator can only terminate if he knows that all the participants know
the final decision. On the other hand, the participant can terminate if he knows
that the coordinator knows that he knows the decision.

\textit{Specification 4a: The coordinator $c$ can always succeed to terminate.}
$$
F ~ (K_{c}  (\hat{K}_{i} (\verb+decision+)))
$$
\textit{Specification 4b: The participants can always succeed to terminate.}
$$
F ~ (K_{i} (K_{c} (\hat{K}_{i}(\verb+decision+))))
$$

\section{The Protocol in The Byzantine-fault Context} \label{Byzantine} \label{sec: Byzantine}

In this context we allow the coordinator to cheat in a variety of ways as shown in the $2PC_{c}$ program. 
We assume that communication is synchronous and reliable.

First note that in order to implement the knowledge-based programs $2PC_{c}$ and $2PC_{i}$ we need to replace the knowledge tests given in these programs by standard tests.
Following the methodology described at Section 3 we introduce some variables that we use to implement the knowledge tests in the programs.
We introduce the variables \verb+retrans_cond[i]+ and \verb+stop_cond[i]+ to implement the knowledge-based program of the coordinator.
The variable \verb+retrans_cond[i]+ is used to capture the conditions under which the coordinator
needs to retransmit the decision to participant $i$, which we use to implement the epistemic condition
$(\neg K_{c} (\hat{K}_{i} (\decision)))$.
The variable \verb+stop_cond[i]+ is used to represent the conditions under which $c$ has to stop sending
the decision to $i$ which we use to implement the epistemic condition
($K_{c} (\hat{K}_{i} (\verb+decision+)))$.
Similarly, we introduce the variables \verb+retrans_cond+, and \verb+stop_cond+
to implement the knowledge-based program of the participants. 
The variable \verb+retrans_cond+ is used to represent the condition concerning retransmission 
($\neg K_{i} (K_{c} (\hat{K}_{i} (\decision)))$).
Finally, the variable \verb+stop_cond+ is used to represent the epistemic condition concerning termination 
($K_{i} (K_{c} (\hat{K}_{i}(\verb+decision+) ))$). 
implementations of the programs.

We first discuss the implementation of the knowledge-based program of the coordinator.
Note that there are two knowledge conditions that need to be implemented: 
termination condition $(K_{c} (\hat{K}_{i} (\verb+decision+)))$ and retransmission condition $(\neg K_{c} (\hat{K}_{i} (\decision)))$.
From the analysis of the MCK counter-examples that result from checking repeatedly the equivalence $(K_{c} (\hat{K}_{i} (\verb+decision+)) \dimp \verb+stop_cond[i]+$)
until it holds, we discover two situations under which the condition $(K_{c} (\hat{K}_{i} (\verb+decision+)))$ holds.
Note that we initially assign the value \false~to the variable \verb+stop_cond[i]+
and then proceed monotonically until we discover all situations under which the above equivalence holds.
The first situation we discover is when $c$ receives an acknowledgement message from $i$ after broadcasting an abort/commit message, 
which implies that $i$ has successfully received the decision of the coordinator. 
There is another, less obvious, situation where the epistemic condition $(K_{c} (\hat{K}_{i} (\verb+decision+)))$ may hold in this context, 
that happens when $i$ votes `no'. 
In this case $i$ knows by the abort-validity requirement of the protocol 
that the only possible decision is to abort. However, when $c$ receives $i$'s
vote it knows that $i$ knows the decision, also by the abort-validity requirement.
On the other hand, when we seek an implementation for the epistemic condition concerning
retransmission in the $2PC_{c}$ program, MCK finds no condition under
which the coordinator has to retransmit the decision to the participants since the Byzantine behaviour of the coordinator
that we consider here does not aim to block or delay termination but aims to corrupt the final outcome of the protocol.
The following two predicates capture these situations.
$$
c.\verb+retrans_cond[i]+ := \false ; ~~
c.\verb+stop_cond[i]+ := (\verb+vote[i]+ = \no ~\lor \verb+ack[i]+)
$$

We turn now to discuss the implementation of the knowledge-based program $2PC_{i}$.
We need first to find an appropriate implementation for termination and retransmission conditions.
One situation in which $K_{i} (K_{c} (\hat{K}_{i}(\verb+decision+)))$ holds is when $i$ sends
a No-vote to the coordinator; again this is due to the abort-validity requirement of the protocol. 
Another situation in which termination condition of $i$ holds in this context is when $i$ receives a commit/abort message from the coordinator.
Since we assume that channels are reliable in this context, the coordinator knows once it sends the decision message that it will be delivered successfully,
and once $i$ receives the message it knows that $c$ knows that $i$ knows the decision.
Similar to the implementation of retransmission condition at the coordinator side,
MCK finds no condition under which $i$ needs to retransmit the acknowledgement message to the coordinator since
in this context every sent message will be delivered successfully from the first transmission trail.
$$ 
i.\verb+retrans_cond+ := \false; ~~ i. \verb+stop_cond+:=  (i.\verb+vote+ = \no ~\lor ~ \decision \neq \undecided )
$$
To complete the implementation of the knowledge-based program $2PC_{i}$ in this context, 
we need to find  concrete implementations
to the knowledge conditions $K_{i} (\verb+cheating+_{c})$ and $\neg K_{i} (\verb+cheating+_{c})$. 
We would like to find the appropriate implementation to the test $K_{i} (\verb+cheating+_{c})$
at two different places in the $2PC_{i}$ program  marked with ($1$) and ($2$)
and for the test $\neg K_{i} (\verb+cheating+_{c})$ at the place marked with ($3$).
The implementations of these knowledge tests are given in Table \ref{table:implementations}.
Note that the  implementation of the epistemic condition $K_{i} (\verb+cheating+_{c})$ at $(1)$ can be different
than its implementation at $(2)$ since the knowledge of the agents concerning cheating may evolve when
the agents collaborate with each other and when they use the traps.
However, for a better understanding of the implementations we may note the following situations in the (extended) protocol in which the conditions $K_{i} (\verb+cheating+_{c})$ and $\neg K_{i} (\verb+cheating+_{c})$ hold.

\begin{itemize}

\item The very obvious situation
in which $K_{i}(\verb+cheating+_{c})$ holds is when $i$ sends a No-vote to $c$ and that the final decision
sent by $c$ is commit. In this case $i$ knows that $c$ is cheating since
the final decision contradicts with the abort-validity requirement of the protocol. 
Note that the agents that vote `no' are always  able to decide
correctly and the coordinator will not be able to cheat them due to the abort-validity requirement. 
On the other hand, the agents that vote `yes' will not be able to determine whether the coordinator is cheating 
due to the commit-validity requirement of the protocol.
This explains why cheating detection is not always common knowledge and why it is necessary for the agents
to collaborate with each other to detect cheating.

\item When we allow the agents to use the traps, we find that they are able to detect cases of cheating
that cannot be detected when they operate independently. 
The first situation is when they discover some contradiction in the decisions sent by the coordinator
since the coordinator is supposed to send non-contradictory decisions to the participants.
The second situation happens when the coordinator sends consistent decisions but incorrect in the sense
that they contradict with the validity requirements of the protocol.

\item The implementation of the test $\neg K_{i} (\verb+cheating+_{c})$ is of interest here since we discover (assisted by automation)
several concrete conditions under which the agent cannot determine whether the coordinator is cheating.
The first case in which $\neg K_{i} (\verb+cheating+_{c})$ holds is when $i$'s vote is $\yes$
and that none of the agents announces for cheating (i.e. $\neg \verb+cheatingDetected+$)
and that the environment does not choose to open the run.
Since no one announced for cheating it implies that the coordinator has not sent decisions
that contradict with the abort-validity requirement. 
Also since the vote of $i$ is $\yes$ $i$ cannot then determine by itself whether the coordinator
is cheating whatever the value of the received decision is. 
This is due to the commit-validity requirement of the protocol.
The second situation in which $\neg K_{i} (\verb+cheating+_{c})$ holds is
when the environment instructs the agents to open the run (i.e. to use a trap) and that after opening the run the agents find that 
all the votes are $\yes$ and all the decisions sent by the coordinator
are consistent. In this case the agents will not be able to determine whether the coordinator
is cheating due to the fact that no one knows
the actual vote of the coordinator. Note that there is no need to use the trap and open the run if some agent announces early that there is a cheating
since in this case the agents will not gain any more knowledge from using the trap.


\end{itemize}
\begin{table}
\begin{center}

\begin{tabular}{|c|c|} 
\hline 
The formula & The equivalent predicates \\
\hline
$K_{i} (\verb+cheating+_{c})$  & ~~~~~~~~~~~~ $(i.\verb+vote+ = \no ~ \land ~ \decision[i] = \commit)$  \hfill{$(1)$}\\
\hline
& $ ( (i.\verb+vote+ = \no ~ \land \decision[i] = \commit) \lor$ $\verb+cheatingDetected+$\\
  $K_{i} (\verb+cheating+_{c})$  &  $ ~\lor ~ ( \verb+trap+ \land (\bigvee_{j, k  =2..n} (\verb+vote+[j] = \no  \land  \decision[k] = \commit) $  \\
& $ ~ \lor ~ (\bigvee_{j \neq k, j, k =2..n} (\decision [j] \neq \decision[k]))) )$ \hfill{$(2)$} \\
\hline
 & ( ($i.\verb+vote+ = \yes$ $\land$ $\neg \verb+cheatingDetected+$ $ ~ \land$ $\neg \verb+trap+$) $\lor$ \\
$\neg K_{i} (\verb+cheating+_{c})$  & ~~$(\verb+trap+ \land (\bigwedge_{i \neq j, j=2 ..n} (\verb+vote+[j] = \yes) \land i.\verb+vote+ = \yes) ~ \land $\\ 
&  ~~~~~ $(\bigwedge_{j \neq k, j, k = 2 ..n} (\decision[j] = \decision[k]$)))  \hfill{$(3)$}\\ 
\hline
\end{tabular} 
\end{center}
\caption{The implementations of the epistemic formulas concerning cheating \label{table:implementations}}
\end{table} 

Note that we consider in this paper a speculative version of the knowledge-based program, in which an agent
follows the decision of the coordinator if $\neg K_{i} (\verb+cheating+_{c})$.
One could also study a conservative version of the protocol, where an agent only follows the coordinator's decision if 
$K_{i} (\neg \verb+cheating+_{c})$ (i.e. if $i$ knows that the coordinator is honest). 
The analysis above shows that in this case the agent would commit only if
it knows the actual vote of the coordinator which is not always possible in the Byzantine fault context
where the coordinator might be faulty. 
From this we note that if the agents use the test $K_{i} (\neg \verb+cheating+_{c})$ in their program
then they will never be able to commit any transaction!


The next question of interest is then whether all cases of cheating can be detected
when adapting the above mentioned strategies. 
The answer obtained by model checking is that it does not, where MCK discovers a counter-example when verifying  Specification 3.
The counter-example happens when all participants votes are \textit{`yes'} and the coordinator sends \textit{consistent} but \textit{incorrect} decisions. 
In this case, since no one knows the actual vote of the coordinator, he can easily cheat the participants without being detected and 
the participants will not benefit from using the traps. 
Note that the proposed cheating detection strategies help to mitigate the impact of cheating 
but do not fully eliminate the possibility of cheating as some kinds of cheating remain undetected (see Table \ref{table:results}).   
We leave to future work the problem of verifying other effective strategies for detecting Byzantine faults.

\begin{table}
  \centering
\begin{tabular}{|c|c|c|c|c|c|c|c|} 
\hline 
 The 2PC protocol & \multicolumn{7}{|c|} {Specifications}\\
\hline
The context & 1a &  1b & 2a & 2b & 3  & 4a & 4b\\
\hline
\hline 
The Byzantine-fault & fails & holds  & fails & fails & fails &  holds&  holds\\
\hline
\end{tabular} 
\caption{The verification results of the protocol in the Byzantine context \label{table:results}}
\end{table}

\section{Finding The Shortest/Longest Run of The Protocol}

Efficiency and performance are necessary issues in the design and development process for distributed protocols. 
We describe an approach, using temporal-epistemic
model checking that can be followed to find the longest and shortest runs in which processes can reach their goal state, and demonstrate its application using our case study. To verify the optimal run of the protocol, we use the model checker to verify formulas of the following form

\begin{equation} \label{BCETFormula}
 \phi = X^{n}~ (\neg (K_{c}  (\hat{K}_{i} (\verb+decision+)))).
 \end{equation}
Note that the negation of formula (\ref{BCETFormula}) represents the epistemic condition concerning termination for the controller (see Specification 4a at Section \ref{CorrectnessCond}). So the goal for process $c$ is to reach a state at which the epistemic formula $(K_{c}  (\hat{K}_{i} (\verb+decision+)))$ holds. Recall that the coordinator can terminate only when he knows that all the participants know the final decision. To discover the optimum run of the protocol, 
we start with the above negative temporal-epistemic formula $\phi$ that says that there is no run where $c$ can terminate at time $n$. The user needs to repeatedly verify $\phi$ while doubling the value of $n$ until
the formula fails, and then use binary search to find the first time the formula fails. 
Let us say that the first time $\phi$ fails is when $n = k$, we then say that the optimum run of the protocol takes $k$-rounds.  However, to discover the details of the optimum run, the user needs to model check $\phi$ at $n =k$, 
the model checker then generates a counterexample that leads him to discover the optimum run.
A counter-example to the specification that the MCK returns when the above formula fails is a sequence of states
$r(0),..., r(k)$ from run $r$ where $(I, r(m)) \models \neg (K_{c}  (\hat{K}_{i} (\verb+decision+)))$ for all $m < k$, and 
$(I, r(k)) \not \models  \neg (K_{c} (\hat{K}_{i} (\verb+decision+)))$. 

On the other hand, to discover the least upper bound for termination of the protocol,
we use the model checker to verify formulas of the following form
\begin{equation} \label{WCETFormula}
 \neg \phi =  (X^{n}~ (K_{c}  (\hat{K}_{i} (\verb+decision+)))).
  \end{equation}
Formula (\ref{WCETFormula}) states that at time $n$ in every run of the protocol the controller can terminate.
Again, the user needs to repeatedly verify $\neg \phi$ while increasing the value of $n$ until he finds the first time where the formula holds (if any). 
We remark here that we model agents using perfect recall semantics so that agents do not lose their knowledge over the time. 
Let us say the first time $\neg \phi$ holds is when $n=w$, we then say that the least upper bound
for termination is $w$. Now to discover the run with the least upper bound, the user needs
to model check $\neg \phi$ at $n=(w-1)$, where $(w-1)$ represents the last round at which $\neg \phi$ fails.
The counter-example generated by model checking $\neg \phi$ at $n= (w-1)$ will lead the user to discover the run
at which $c$ cannot terminate at round $(w-1)$ and needs one more round, which represents the
least upper bound for termination.

In Table \ref{tablebcs-wcs} we summarise the lower bound and the least upper bound for termination in each context in terms of the number of
communication rounds and message traffic, where $d$ denotes the number of agents in the environment.
It is interesting to mention that the derived implementations for the two contexts have been verified for $d =  2, 3, 4$ and 5. We conjecture that our present implementation can be shown to work
for all number of agents and it would be interesting to have a proof of this claim: 
this would have to be done manually unless an induction can be found
for the model checking approach.
 
 \begin{table} 
  \centering
  \begin{tabular}{|c|c|c|c|c|} 
\hline 
The 2PC protocol & \multicolumn{2}{|c|} {The Shortest run} &  \multicolumn{2}{|c|} {The Longest run} \\
\hline
The context & \verb+#+of rounds ~ &  \verb+#+of messages~ & \verb+#+of rounds ~&  \verb+#+of messages \\
\hline
The Byzantine-fault context &  1  &	 $(d-1)$	&	3	& $3(d-1)$ \\
\hline
\end{tabular} 
\caption{The best and worst case scenarios of the protocol in the Byzantine context \label{tablebcs-wcs}}
\end{table}

\section{Conclusion and Future Work}

In this paper, we have used the knowledge-based approach and a methodology based on model checking to derive a family of protocols
for the distributed transaction commit problem in a Byzantine faults context.
The problem of analysing the correctness of distributed algorithms specifically in the presence of (Byzantine) faults in an \textit{automated fashion} is very important and has been largely ignored by both the formal methods and distributed algorithms communities. In future, we aim to verify some other mechanisms for detecting Byzantine faults in particular when replicating the transaction coordinator  and employing some cryptographic techniques.  There are some protocols such as various distributed consensus protocols under Byzantine faults that are extremely tricky to reason about, and thus
model checking tools for knowledge-based programs could potentially be
used to optimize their behaviour. 

\bibliographystyle{abbrv}
\bibliography{references}

\end{document}